\documentclass{aa}  
\usepackage{graphicx}
\usepackage{xcolor}

\usepackage{caption}
\definecolor{links}{rgb}{0, 0, 255}
\usepackage[colorlinks=true,allcolors=links]{hyperref}
\usepackage[varg]{txfonts}
\newcommand{\mpchi}{\,h^{-1}{\rm {Mpc}}}

\newcommand{\msun}{M_{\sun}}
\newcommand{\msunh}{h^{-1}M_{\sun}}

\newcommand{\hi}{H{~\sc i}\xspace}
\newcommand{\hj}{\text{\hi}}

\begin{document}

\title{\textsc{NeutralUniverseMachine}: Predictions of \hi\ gas in Different Theoretical Models}

\author{Lingkun Wen\inst{\ref{inst1},\ref{inst2}}
\and Hong Guo\inst{\ref{inst1}}\fnmsep\thanks{Corresponding Author}
\and Wenlin Ma\inst{\ref{inst1},\ref{inst2}} 
\and Lizhi Xie\inst{\ref{inst3}}
\and Gabriella De Lucia\inst{\ref{inst4}}
\and Fabio Fontanot\inst{\ref{inst4},\ref{inst5}} 
\and Michaela Hirschmann\inst{\ref{inst6}}      
}

\institute{
Shanghai Astronomical Observatory, Chinese Academy of Sciences, Shanghai 200030, China.\label{inst1} \email{guohong@shao.ac.cn}
\and University of Chinese Academy of Sciences, Beijing 100049, China.\label{inst2}\email{wenlingkun@shao.ac.cn}
\and Tianjin Normal University, Binshuixidao 393, 300387, Tianjin, China \label{inst3}
\and INAF-Astronomical Observatory of Trieste,   {Via G.B. Tiepolo, 11 I-34143 Trieste, Italy}\label{inst4}
\and IFPU - Institute for Fundamental Physics of the Universe, via Beirut 2, I-34151, Trieste, Italy\label{inst5}
\and DARK, Niels Bohr Institute, University of Copenhagen, Lyngbyvej 2, DK-2100 Copenhagen, Denmark\label{inst6}
}

\abstract
  % context heading (optional)
{We investigate the distribution and evolution of \hi gas in different theoretical models,   {including hydro-dynamical simulations (Illustris-TNG and SIMBA), semi-analytic models (GAEA), and the empirical models (\textsc{NeutralUniverseMachine}; NUM). By comparing} model predictions for the \hi mass function (HIMF), \hi-halo and \hi-stellar mass relations, conditional \hi mass function (CHIMF) and the halo occupation distribution (HOD) of \hi-selected galaxies, we find that all models show reasonable agreement with the observed HIMF at $z\sim0$, but the   {differences} become much larger at higher redshifts of $z=1$ and $z=2$. The HIMF of NUM shows remarkable agreement with the observation at $z=1$,   {whereas} other models predict much lower amplitudes of HIMF at the high-mass end. Comparisons of CHIMF distributions indicate that the HIMF is dominated by halos of $10<\log(M_{\rm vir}/\msun)<11$ and $11<\log(M_{\rm vir}/\msun)<12$ at the low- and high-mass ends, respectively. From the \hi HODs of central galaxies, we find that TNG100 overpredicts the number of central galaxies with high $M_\hj$ in massive halos and GAEA shows a very strong depletion of \hi gas in quenched centrals of massive halos. The main cause of the   {differences} is the AGN feedback mechanisms implemented in different models. 
}

\maketitle
   
\section{Introduction}\label{sec_Intro} 
% Galaxies play a fundamental role in the large-scale structure of the universe, whereas the cold gas serves as the primary fuel of star formation.
  {Cold gas, composed mainly of its atomic and molecular elements, with atomic hydrogen (\hi) playing a major role, acts as the essential fuel for star formation. Consequently, investigating the \hi distribution is vital to understand the processes of galaxy formation and its interaction with dark matter halos, and it can also facilitate in comprehending the large-scale structure of galaxies.} The distribution of the \hi gas in the local universe   {has been} accurately mapped through \hi 21~cm surveys, such as the Arecibo Fast Legacy ALFA Survey \citep[ALFALFA;][]{Giovanelli_2005, Haynes_2011},  MeerKAT \hi Intensity Mapping \citep{santos_2015}, and FAST All Sky \hi Survey \citep[FASHI;][]{zhang_fast_2024}. However, detecting individual galaxies becomes increasingly challenging at higher redshifts because of weak signals and radio frequency interference. To overcome this, techniques such as \hi spectra stacking and intensity mapping have been used to estimate the cosmic \hi density for redshifts up to $z<1$ \citep[e.g.,][]{Lah_2007, Delhaize_2013, Masui_2013, Rhee_2013,Rhee_2018, Kanekar_2016, Bera_2019, Guo_2020}. 

  {Currently,} most of the \hi observations beyond the local universe are concentrated around $z\sim 0.37$, with only a few observations at $z\sim 1$, and even fewer at higher redshifts. However, simulations can provide an opportunity to understand the underlying physical processes. Various models have been proposed to explain the observations, including hydrodynamical simulations, semi-analytical models (SAMs), and empirical models. Depending on the adopted subgrid physics in modelling the galaxy formation and evolution, hydrodynamical simulations can reasonably replicate the observations in the stellar components such as the stellar mass function and star formation rate. However, the \hi and H$_2$ gas properties in these simulations   {could} vary significantly. As shown in \cite{Dave2020}, the hydrodynamical simulations of SIMBA \citep{Dave2019}, IllustrisTNG \citep{Nelson_2019}, and EAGLE \citep{Schaye2014,Crain_2015} exhibit strong   {differences} in their \hi properties, particularly at higher redshifts. 

One of the key challenges in hydrodynamical simulations is the extensive resource requirement needed to conduct large-volume high-resolution simulations.   {It is also extremely difficult to model the transition between atomic and molecular hydrogen from the first principle. Therefore, it is generally treated using a post-processing framework} \citep{Duffy2012, Lagos2015,Diemer_2018}, similar to the approach used in SAMs \citep[see e.g.,][]{Blitz_2006,Fu2013,Xie_2017}. By leveraging halo merger trees from high-resolution $N$-body simulations, SAMs can effectively   {model} the cold-gas components of galaxies over significantly larger volumes compared to hydrodynamical simulations. The model parameters are often fine-tuned using the observed cold-gas properties at z=0. However, the varied approaches of galaxy formation employed in these models can result in a notable   {differences} in other statistics that are typically uncalibrated. For example, \cite{Baugh2019} identified substantial variations in the \hi -halo mass relations between different models, especially for halos with $M_{\rm vir}>10^{11.5}\msun$ where the AGN feedback becomes a significant factor \citep[see also][]{Chauhan2020,Spinelli_2020}. 

% In principle, it is difficult for SAMs to fit all available stellar and cold-gas observations by exploring the huge model parameter space, given the complicated galaxy formation models 
  {In practice, although the parameter space is large, it is difficult for SAMs to fit well all available stellar and cold-gas observational measurements}
\citep[see e.g.][]{Henriques2015}. Using simple functional forms without involving detailed baryon physics, empirical models become another popular way to   {statistically link} the properties of galaxies and their halo environment \citep[see e.g.,][]{Behroozi2013,Behroozi_2019}. Recently, \cite{guo2023NUM} proposed a new empirical model, \textsc{NeutralUniverseMachine}, which is capable of accurately describing most observed cold gas properties, including the \hi and H$_2$ mass functions, the \hi -halo and \hi -stellar mass relations \citep{Guo_2020, Guo_2021}, H$_2$-stellar mass relation \citep{Saintonge2017}, and the redshift evolution of the cosmic \hi and H$_2$ densities \citep{Walter2020}. It can be used to further investigate the distribution and evolution of cold gas \citep[see e.g.,][]{Ma_2024}. 

  {As we will show in the following sections, the comparisons of different model predictions provide more insight into the physical mechanisms that affect the cold-gas distributions.} In this study, we will compare the predictions of \textsc{NeutralUniverseMachine} with a set of different hydrodynamical and SAMs. In particular, we will explore the model predictions of the \hi-halo mass relation, \hi-stellar mass relation, conditional \hi mass function, and the halo occupation distribution of \hi-selected galaxies from $z=0$ to $z=2$.

This paper is structured as follows. In Section ~\ref{sec_data}, we introduce the simulations and data that we used. We present our main results in Section ~\ref{sec_results}. In Sections ~\ref{sec_dicussion} and ~\ref{sec_conclusion}, we provide our discussions and conclusions.
	
\section{Simulations}\label{sec_data}
\subsection{The IllustrisTNG Simulation}\label{subsec_TNG}
The IllustrisTNG project consists of a set of magnetohydrodynamical simulations of different cosmological volumes \citep{Marinacci2018,Naiman2018,Nelson2018,Nelson_2019,Pillepich2018,Springel2018}, run with the Arepo code \citep{Springel2010}. We focus only on the TNG100-1 simulation (referred to as TNG hereafter), which has a box size of $75\mpchi$ on each side and the cosmological parameters of $\Omega_{\rm m}=0.3089$, $\Omega_{\rm \lambda}=0.6911$, $\Omega_{\rm b}=0.0486$ and $h=0.6774$. The resolution of the dark matter mass is $m_{\rm DM}=7.5\times10^{6}\msun$ and the average mass of the gas cells is approximately $m_{\rm gas}=1.4 \times 10^{6}\msun$.  Dark matter halos are identified using a standard FoF algorithm with a linking length of b = 0.2, while subhalos are identified in FoF groups using the SUBFIND algorithm \citep{Springel_2001}. 

The \hi gas mass is calculated with the post-processing framework presented in \cite{Diemer_2018}, where five different models have been used to model the atomic-to-molecular transition. The \hi and $\rm H_2$ masses are measured within the whole subhalo containing the corresponding galaxy. In this work, we only use the output of the 'K13' model \citep{Krumholz_2013} with projected quantities in \cite{Diemer_2018}.   {As demonstrated in the Appendix of \cite{Ma2022}, the impact of various transition models on \hi gas scaling relations is negligible. Thus, adopting the K13 model is for better comparisons with previous studies.}

This model operates under the assumption of a threshold density ($n_{\rm H} = 0.106 \, \text{cm}^{-3}$, \citealt{Diemer_2018}) above which star formation occurs. While the initial hydrogen fraction is set at 0.76, it decreases as the gas becomes enriched with helium and metals.   {The model also consider the interplay between $\rm H_2$ recombination and UV dissociation in molecular clouds.}

\subsection{The SIMBA Simulation}\label{subsec_SIMBA}
The SIMBA simulations \citep{Dave2019} are run with the GIZMO code \citep{Hopkins2015} which offers a solution for cosmological gravity and hydrodynamics. In this study, we only use the SIMBA flagship run (m100n1024) with a box size of $100\mpchi$ on each side and the cosmological parameters are similar to those in TNG, with $\Omega_{\rm m}=0.3$, $\Omega_{\lambda}=0.7$, $\Omega_{\rm b}=0.048$, and $h=0.68$. The resolutions for dark matter particles and gas cells are $9.6 \times10^{7}\msun$ and $1.82\times 10^{7}\msun$, respectively. In SIMBA, galaxies are identified through a 6D friends-of-friends (FOF) finder, with a linking length equivalent to 0.0056 times the mean interparticle distance. Furthermore, halos are identified using a 3D FOF finder with a link length parameter of 0.2.

SIMBA adopts the H$_2$-based star formation law of \cite{Schmidt1959}, which takes into account the local metallicity and gas column density \citep{Krumholz2011}. We refer the readers to \cite{Dave2020} for more details. The star formation rate (SFR) is then calculated using the equation: 
\begin{equation}
    {\rm SFR}= \varepsilon_{\ast} f_{\rm H_2}\rho/t_{\rm dyn}
\end{equation}
where $\varepsilon_{\ast}=0.02$ \citep{Kennicutt1998}, $f_{\rm H_2}$ is the H$_2$ fraction of each gas element, $\rho$ is the gas density, and the $t_{\rm dyn}$ is the dynamical time scale. The \hi fraction of each gas element is determined   {by} the prescription of \cite{Rahmati2013}.   {The \hi and H$_{2}$ proportions for each particle are obtained directly from the simulation, with no further post-processing.} Measurements of the \hi mass function and the \hi-stellar mass relation have been presented in \cite{Dave2020}.

\subsection{GAEA Model}\label{subsec_GAEA}
The GAlaxy Evolution and Assembly (GAEA) model is a semi-analytical model built on the work of \cite{DeLucia_2007}. In this study, we adopt the latest GAEA model (GAEA2023) as described in \cite{DeLucia_2024}. It is applied in two simulations of different resolutions, and we choose the one with higher resolution, based on the MillenniumII $N$-body simulation \citep{Boylan2009}. The simulation consists of $2160^3$ dark matter particles with a box size of $100\mpchi$ on a side, assuming a WMAP1 cosmology of $\Omega_\Lambda = 0.75, \Omega_{\rm m} = 0.25, \Omega_{\rm b} = 0.045, \sigma_8 = 0.9, h = 0.73$. The dark matter particle mass of the simulation is $6.9\times 10^6 \msunh$. 

% In comparison to the initial version, 
  {GAEA2023 incorporates} an in-depth analysis of non-instantaneous baryon recycling \citep{DeLucia_2014},   {a parameterisation   {partially} based on the results of high-resolution hydrodynamical simulations of stellar feedback \citep{Hirschmann_2016}, a specific division of the cold gas reservoir into its atomic and molecular components} \citep{Xie_2017}, a meticulous tracking of angular momentum transfer \citep{Xie_2020}, an enhanced model for AGN feedback \citep{Fontanot_2020}, and a novel approach to the ram-pressure stripping of gas from satellite galaxies \citep{Xie_2020,Xie2024}. 

In GAEA2023, the star formation rate   {depends} on the quantity of cold gas, defined as the total gas in galaxies with temperatures below $10^4 \rm K$. Within this, it is assumed that 26 percent of the cold gas comprises helium, dust, and ionised gas at all redshifts, with the remaining gas categorised as \hi and $\rm H_2$.   {GAEA2023 uses the \hi-H$_2$ transition model of \citep{Blitz_2006}, in which the $\rm H_2$/\hi ratio depends on the mid-plane pressure of the gas disk. The model parameters are calibrated to match the observed \hi mass function, $\rm H_2$ mass function, and the galaxy stellar mass function at $z = 0$.}

% Then the star formation rate can be calculated as\citep{Xie_2017}:
% \begin{equation}
%     \Sigma_{\rm sfr} = \nu_{\rm sfr} \Sigma_{\rm H_2}
% \end{equation}
% where $\Sigma_\rm sfr/H_2$ denotes the surface density, and $\nu_{\rm sfr}$ represents the effectiveness of transforming $\rm H_2$ into stars, which varies across different star formation theories.

\subsection{NeutralUniverseMachine Model}\label{subsec_NUM}
The \textsc{NeutralUniverseMachine}\footnote{\url{https://halos.as.arizona.edu/UniverseMachine/DR1/Gas_Masses_NeutralUniverseMachine/}} (hereafter NUM) model is an empirical model to accurately describe the evolution of \hi and H$_2$ gas in the halos in $0<z<6$ \citep{guo2023NUM}. It is based on the UniverseMachine catalogue of \cite{Behroozi_2019} using the Bolshoi-Planck N-body simulation \citep{Klypin2016}. The simulation has a box size of $250\mpchi$ on each side   {(though the box size of NUM is larger than others, the volume effect is negligible)} and a dark matter particle mass resolution of $2.3\times10^8\msun$. The cosmological parameters are $ \Omega_{\rm m} = 0.307$, $h = 0.678$, $\Omega_{\rm b} = 0.048 $, and $\sigma_8 = 0.823$, consistent with those of TNG. The dark matter halos and subhalos are identified with the ROCKSTAR halo finder \citep{Behroozi_2012}. 

  {The NUM model   {determines} the \hi mass of a galaxy ($M_{\hj}$) based on its halo mass ($M_{\rm vir}$), halo formation time ($z_{\rm form}$), SFR and $z$, as follows:}
\begin{align}\label{eq_numhi}
    M_\hj =&~ \frac{\kappa M_{\rm vir}}{\mu^{-\alpha} + \mu^\beta} \left( \frac{1+z}{1+z_{\rm form}} \right)^\gamma
    \left( \frac{\rm SFR}{\rm SFR_{\rm MS,obs}} \right)^\lambda\\
    \mu =&~ M_{\rm vir}/M_{\rm crit}\\
    \log \kappa =&~ \kappa_0 + \kappa_1 z + \kappa_2 z^2\\
    \log M_{\rm crit} =&~ M_0 + M_1 z + M_2 z^2
\end{align}
  {where SFR is measured in units of $M_\odot/\rm yr$ and $\rm SFR_{\rm MS,obs}$ represents the best-fit star formation main sequence (SFMS) presented in \cite{Behroozi_2019}, with $\kappa_0$, $\kappa_1$, $\kappa_2$, $M_0$, $M_1$, $M_2$, $\alpha$, $\beta$, $\gamma$, and $\lambda$ being the free model parameters.}

  {NUM adopted the $\rm H_2$ mass model defined in \cite{Tacconi2020}:}
\begin{align}\label{eq_numh2}
    M_{\rm H_2} = &~ \zeta M_\ast^\mu \left( \frac{\rm SFR}{{\rm SFR}_{\rm MS,obs}}\right)^\eta\\
    \log \zeta =&~ \zeta_0 + \zeta_1\ln(1+z) + \zeta_2[\ln(1+z)]^2
\end{align}
  { where $\zeta_0, \zeta_1, \zeta_2, \mu, $  and $\eta$ are free model parameters.}

The NUM model   {calibrates the free parameters by fitting to}  the observed mass functions of \hi and H$_2$, the molecular-to-atomic ratio, the \hi-halo and \hi-stellar mass relations, the \hi clustering measurements at $z\sim 0$, and the \hi-stellar mass relations at higher redshifts. Furthermore, it accurately describes the evolution of cosmic gas densities $\rho_{\text{\hi}}$ and $\rho_{\rm H_2}$ in the redshift range of $0<z<6$ \citep{guo2023NUM}. This empirical model offers the advantage of accurately describing the observed gas properties without relying on complex baryon physics. Furthermore, it allows for precise constraints on the cold gas content through various observations. Consequently, it can be used to make predictions for different relationships, which can then be compared with future observations. It is shown in \cite{Ma_2024} that the NUM model shows very good agreement with the observed \hi-halo and \hi-stellar mass relations in different large-scale environments quantified by the distances to filaments. 

% \subsection{Star-forming and Quenched Galaxies} \label{sec_sample}

% In the case of galaxies having similar stellar masses, it is observed that star-forming galaxies generally exhibit significantly higher \hi masses compared to their quenched counterparts \citep{Guo_2021}. Quenched galaxies are particularly valuable for studying the diverse feedback mechanisms in various galaxy formation models. Hence, it is crucial to compare the \hi characteristics of star-forming and quenched galaxies. In this study, we present our criteria for distinguishing between these two populations. 

\begin{figure*}
    \centering
    \includegraphics[width=0.65\textwidth]{ 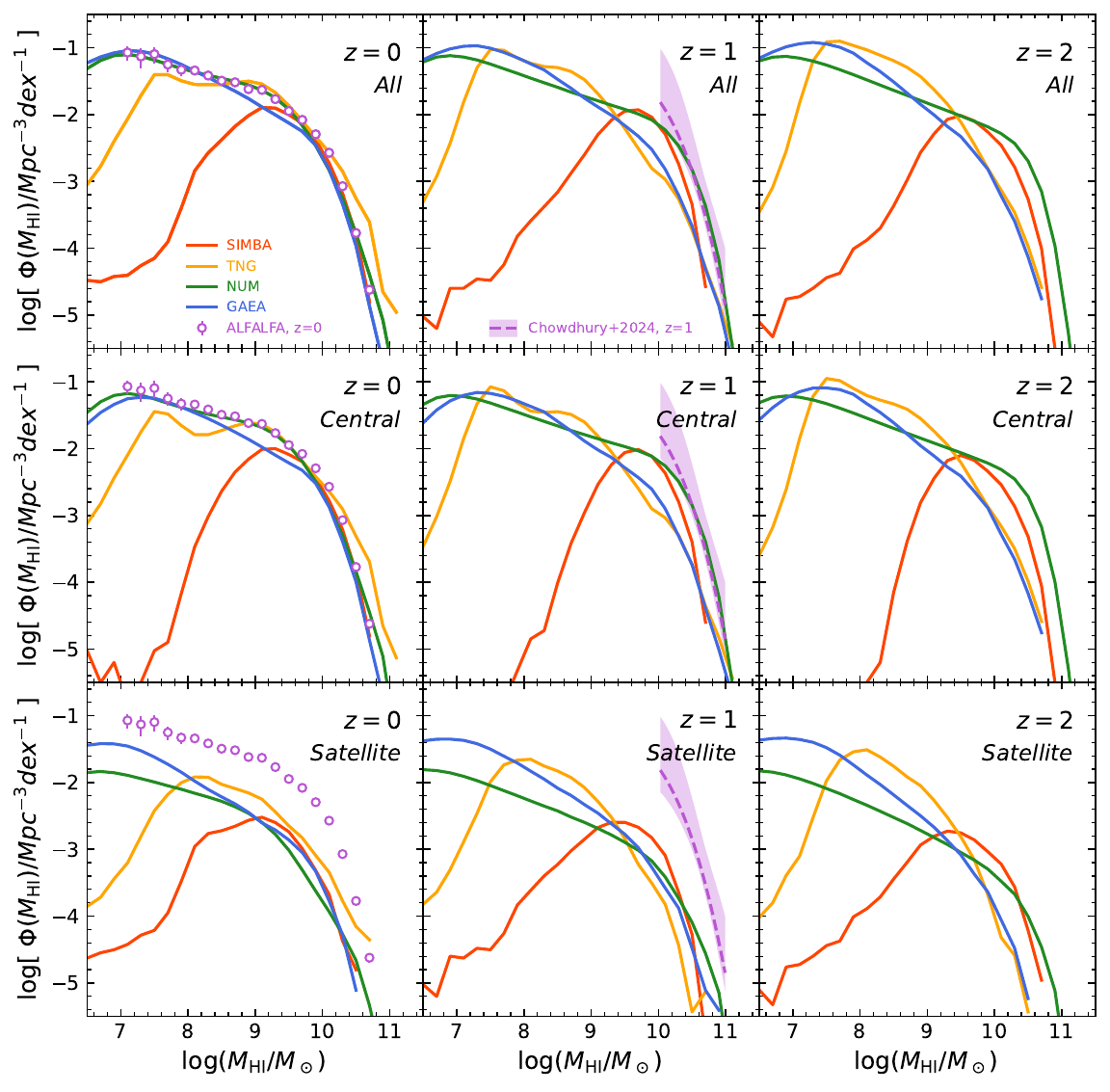}
    \caption{Comparisons between the HIMF of the different simulations and the observations. Solid lines represent the models, and the observations for all galaxies at z$\sim$0 are shown in purple open dots. The results of all SFGs at z$\sim$1 from \cite{Chowdhury2024} are plotted in dashed lines, and the 1$\sigma$ errors are shown as the shaded region.   {The high-mass range of the HIMF is largely contributed by SFGs, therefore the SFG results at z$\sim$1 would be similar to those for the entire galaxy population.} The top row represents the HIMF for all galaxies, and the middle and bottom rows show the HIMF for central and satellite galaxies, respectively. Different columns represent different redshifts. }\label{fig_HIMF_all}
\end{figure*}

\section{Results}\label{sec_results}

\subsection{\hi Mass Function}\label{subsec_HIMF}
The \hi mass function (HIMF), represented by $\phi(M_{\text{\hi}})$, is a measure of the average number densities of galaxies in specific \hi mass ranges. It provides valuable information about the distribution of cold gas and its variation over time clearly indicates the evolution of gas. The HIMF in the local universe was determined using the final release of ALFALFA data \citep{Jones2018}. In this study, we used the HIMF measurements of ALFALFA from \cite{guo2023NUM}   {(see their Table.~2) with a completeness threshold of 90\%}, which correct for the incompleteness effect at the low-mass end of the original measurements by \cite{Jones2018}.   {We note that the observed HIMF of ALFALFA is for HI targets with optical counterparts with the minimum stellar mass at around $10^6\msun$ \citep{Huang2012}. Therefore, we apply the same stellar mass resolution limits of $M_\ast > 10^6 M_\odot$ in all theoretical models for fair comparisons.}

Figure \ref{fig_HIMF_all} illustrates the comparisons of the HIMF measurements from various models, denoted by lines of different colours. To facilitate comparison, the observed HIMF measurements at $z=0$ from \cite{guo2023NUM} are plotted in circles with error bars. The left, middle and right panels show the model predictions for $z=0$, $z=1$, and $z=2$, respectively. Furthermore, the top, middle, and bottom rows correspond to all, central, and satellite galaxies, respectively.

  {The results for all star-forming galaxies (SFGs) at $z \sim 1$ from \cite{Chowdhury2024} are depicted with dashed lines, while the shaded regions indicate the 1$\sigma$ errors. Since the HIMF is primarily contributed by the SFGs, the resulting HIMF of the SFGs would be mostly consistent with that of the entire galaxy sample). \cite{Chowdhury2024} combine the B-band luminosity function, $\Phi (M_{\rm B})$, with the $M_\hj$-$M_{\rm B}$ measurements at $z \sim 1$ to infer the HIMF. They estimated that the measurement error in the $M_\hj$-$M_{\rm B}$ relation is around $0.26 ~\rm dex$. } 

\begin{figure}
    \centering
    \includegraphics[width=0.45\textwidth]{ 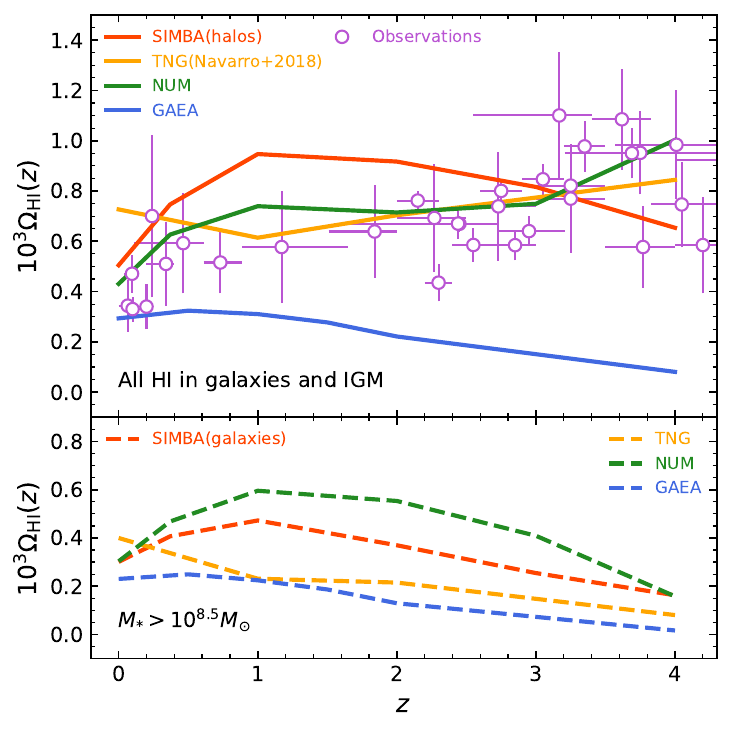}
    \caption{Comparisons between \hi fraction, $\Omega_{\text{\hi}}(z)$, in different simulations (solid lines) and the observations of \cite{Walter2020} (open circles).   {The top panel show all \hi in the galaxies and IGM with no selection. } We refer the readers to Fig.~4 of \cite{guo2023NUM} for more details about the observations.   {The bottom panel shows $\Omega_\hj$ of different theoretical models with a mass threshold of $M_\ast > 10^{8.5}M_\odot$ (dashed lines).} The labels are the same as in Figure \ref{fig_HIMF_all}.} 
    % The gas masses at $z > 6$ ($z > 5$ for \hi) are not constrained in the NUM model and that for we only plot $\Omega_{\text{\hi}}(z)\in (0,4)$.
    \label{fig_OmegaHI}
\end{figure}

The top left panel clearly shows that most models reasonably represent the observed HIMF at $z=0$. The TNG model exhibits a slight excess of \hi-rich galaxies at the high-mass end. SIMBA shows a decreasing trend for $M_\hj<10^9\msun$ due to the limited mass resolution \citep{Dave2020}. Therefore, we will only focus on   {galaxies with} $M_\hj>10^9\msun$ for SIMBA. The evolution trends of the HIMF vary in different models. In GAEA and TNG,   {the massive end of HIMF increases} as the redshift decreases, while   {low-mass end of HIMF} decreases over time.   {we confirm that the HIMFs of GAEA at the massive end are not affected by the small simulation volume of MillenniumII. Adopting the GAEA implementation in the Millennium Simulation with a larger box size of 500 $\rm Mpc/h$ produces consistent HIMFs at the massive end.} In contrast, there is only weak evolution at the low-mass end in NUM, while the   {massive end} quickly decreases towards $z=0$,   {resulting from the descending trend of $\Omega_{\hj}$ shown in Figure~\ref{fig_OmegaHI}}. The evolution of the HIMF in SIMBA shows a relatively minor change, with a slightly higher   {HIMF in massive end} at higher redshifts.

  {However, we note that the observed measurements of HIMF could also suffer from systematic effects, such as the cosmic variance effect at the massive end. The HIMF of \cite{Chowdhury2024} at $z\sim1$ is based on the measurements in the small sky area of the DEEP2 survey, which could be potentially affected by the cosmic variance effect. The HIMF of the ALFALFA survey at $z\sim0$ is based on a much more reliable sample covering around 7000 square degrees. Moreover, the typical measurement error of \hi mass in ALFALFA is around 0.05~dex, which makes the effect of Eddington bias minimal. }

Most of the existing blind surveys in \hi are limited to low redshifts of $z<0.5$. As a result, it is challenging to differentiate between various model forecasts regarding the evolution of HIMF. However, recent measurements of HIMF beyond the local universe tend to favour much more \hi-rich galaxies at higher redshifts \citep{Xi2021,Chowdhury2024}. In the middle panels of Figure~\ref{fig_HIMF_all}, we show the HIMF measurement of \cite{Chowdhury2024} using Giant Metrewave Radio
Telescope (GMRT) observations at $z\sim1$ for SFGs with $M_\hj>10^{10}\msun$ as the dashed line, and the corresponding errors are shown as the shaded region. The NUM model is in very good agreement with the measurement of \cite{Chowdhury2024},   {reproducing} the strong evolution of the HIMF at the massive end. 

When analysing the contributions to the HIMF by distinguishing between central galaxies (middle panels) and satellite galaxies (bottom panels), it becomes apparent that central galaxies are the primary contributors to the HIMF across all redshifts from 0 to 2, and this pattern is consistent across all models. This is mainly due to the stellar mass functions being higher for central galaxies than for satellites. Gas depletion in satellite galaxies within a high-density halo environment \citep{Brown2017,Stevens2019} only has a secondary impact. Similarly to the top panels, the   {differences} among different models become more pronounced at higher redshifts.

\subsection{Cosmic H~$\scriptsize{I}$ abundance  $\Omega_\hj$}
% It is intriguing to investigate the evolution of cosmic \hi abundance, 
  {The $\Omega_\hj$ (i.e., the total \hi density in the universe), can be estimated by directly integrating the HIMF in observation} \citep[see e.g.,][]{Jones2018} or from the \hi column density measurements of damped Lyman-$\alpha$ systems \citep[see e.g.,][]{Wolfe_2005}.   {In simulations, we simply calculate the \hi fraction as: $\Omega_\hj \equiv \rho_\hj/\rho_{\rm c}$, where $\rho_\hj$ is the density of \hi mass, $\rho_{\rm c} $ is the critical density at z = 0.} In Figure \ref{fig_OmegaHI}, we compare the various model predictions with the collection of observed measurements (open circles) from \cite{Walter2020}.
  {The top panel illustrates the results for all \hi present in galaxies and the intergalactic medium (IGM), represented by solid lines. The bottom panel displays the data after imposing a resolution threshold of $M_\ast > 10^{8.5} M_\odot$, shown in dashed lines.}   {In the top panel of Figure \ref{fig_OmegaHI}, we observe that the NUM model} fits the observations reasonably well   {by accounting for orphan galaxies (i.e., subhalos experiencing tidal stripping resulting in masses below the simulation resolution) \citep{Behroozi_2019, guo2023NUM}. Consequently, the NUM model will serve as a reference for comparisons.} 

  {
\cite{Navarro_2018} conducted an extensive analysis of \hi in TNG, illustrating $\Omega_\hj$ in their Fig.~2. Likewise, \cite{Diemer_2019} investigated $\Omega_\hj$ in TNG\footnote{Corrected measurements of their result are available at \url{http://www.benediktdiemer.com/data/hi-h2-in-illustris/}}, finding a 30\% lower $\Omega_\hj$ than that of \cite{Navarro_2018}.} As shown in \cite{Navarro_2018}, the contribution of IGM to $\Omega_{\hj}$ in TNG is less than 5\% at $z<2$ and around 5\%\text{-}15\% at $2<z<4$, which cannot explain the significant difference.   {We note that the post-processing of \hi and H$_2$ in \cite{Diemer_2019} is only applied for galaxies with $M_\ast>2\times10^8\msun$ or $M_\hj>2\times10^8\msun$, but it is applied for all gas cells in \cite{Navarro_2018}. The mass limit in \cite{Diemer_2019} leads to the much smaller $\Omega_\hj$. The different post-processing models only have a minor effect \citep{Diemer_2019}. Since the observational measurements of $\Omega_\hj$ include the \hi in galaxies and IGM, we employ the results of \cite{Navarro_2018} in the top panel (solid orange line) and only show the TNG results of \cite{Diemer_2019} in the bottom panel. Except for the excess at $z<0.5$, $\Omega_\hj$ in TNG closely matches those from observations and NUM up to high redshifts.}

  {In SIMBA, all gas particles in a halo are assigned to the galaxy with the highest $M_{\rm baryon}/R^2$, where $M_{\rm baryon}$ is the baryonic mass of the galaxy, and $R$ is the distance of a given gas particle to the centre of the galaxy. This method identifies \hi as ISM even in low-density areas ($n_{\rm H} < 0.13 ~\rm cm^{-2}$, \citealt{Dave2020}). However, in the absence of a galaxy that surpasses the simulation resolution threshold ($M_\ast \sim 10^{8.4} \msun$), the gas particles in a halo would not be associated with any galaxy. The resolution effect results in a substantial amount of \hi in the IGM for low-mass halos. The total predicted $\Omega_\hj$ from the SIMBA model (solid red line) aligns with observations for redshift ranges $z<0.5$ and $z>3$, but is substantially overestimated for $0.5<z<3$.}

  {For GAEA, $\Omega_\hj$ is significantly underestimated at most redshifts and only shows agreement with observations at $z\sim0$. As discussed in \cite{Yates_2021}, most semi-analytical models, including GAEA, are limited by the inadequate modelling of \hi in IGM and circumgalactic medium (CGM), as well as in sub-resolution halos. \cite{Gioia_2020} tried to overcome the resolution effect by using a halo occupation distribution method to improve the model of \hi in low-mass halos, but it did not improve the model predictions at $z>2$. As noted by \cite{Spinelli_2020} and \cite{Yates_2021}, the correction to the sub-resolution halos is not enough to resolve the differences.} 

%  {TNG tends to overestimate \hi at $z=0$, mainly due to a surplus at the high-mass end ($M_\hj > 10^{10} M_\odot$) in HIMF. Meanwhile, GAEA tends to underestimate the HIMF within the mass range $M_\hj \sim 10^8 \text{-} 10^{10} M_\odot$ at $z = 0$, and also for $M_\hj > 10^{10} M_\odot$ when $z > 1$, leading to notable differences from observations (see Fig.~\ref{fig_HIMF_all}). Discrepancies observed in both TNG and GAEA highlight the need for precise HIMF modelling, particularly at high \hi mass ($M_\hj > 10^{8} M_\odot$). }

  {To compare the different models for only well-resolved galaxies, we show in the bottom panel of Figure~\ref{fig_OmegaHI} the results after imposing the stellar mass threshold of $M_\ast > 10^{8.5}\msun$. It is interesting that the differences among the theoretical models become smaller because of the removal of a large portion of \hi in low-resolution halos. Compared to the top panel, $\Omega_\hj$ of NUM only decreases by around 20\%--30\% for $z<2$, but decreases sharply at higher redshifts. It emphasises the dominant contribution of \hi in low-mass galaxies at high redshifts. Accurate modelling of \hi in these galaxies is key to understanding the evolution of $\Omega_\hj$.} 

  {Unlike the top panel, $\Omega_\hj$ of TNG for these well-resolved galaxies is becoming much smaller than that of NUM for $z > 0.5$. Since the contribution of IGM to $\Omega_\hj$ is still small at these redshifts \citep{Navarro_2018}, it suggests that TNG might underestimate the \hi mass in massive galaxies at high redshifts, as will be shown in the following sections. The resolution effect is more severe for SIMBA, with $\Omega_\hj$ decreasing significantly for galaxies above $10^{8.5}\msun$. The predictions of SIMBA get closer to that of NUM for these well-resolved galaxies. Compared to the top panel, $\Omega_\hj$ of GAEA exhibits a minimal shape change. It confirms the above conclusion that resolution effect is not the main culprit of the differences.}

\begin{figure*}
    \centering
    \includegraphics[width=0.73\textwidth]{ 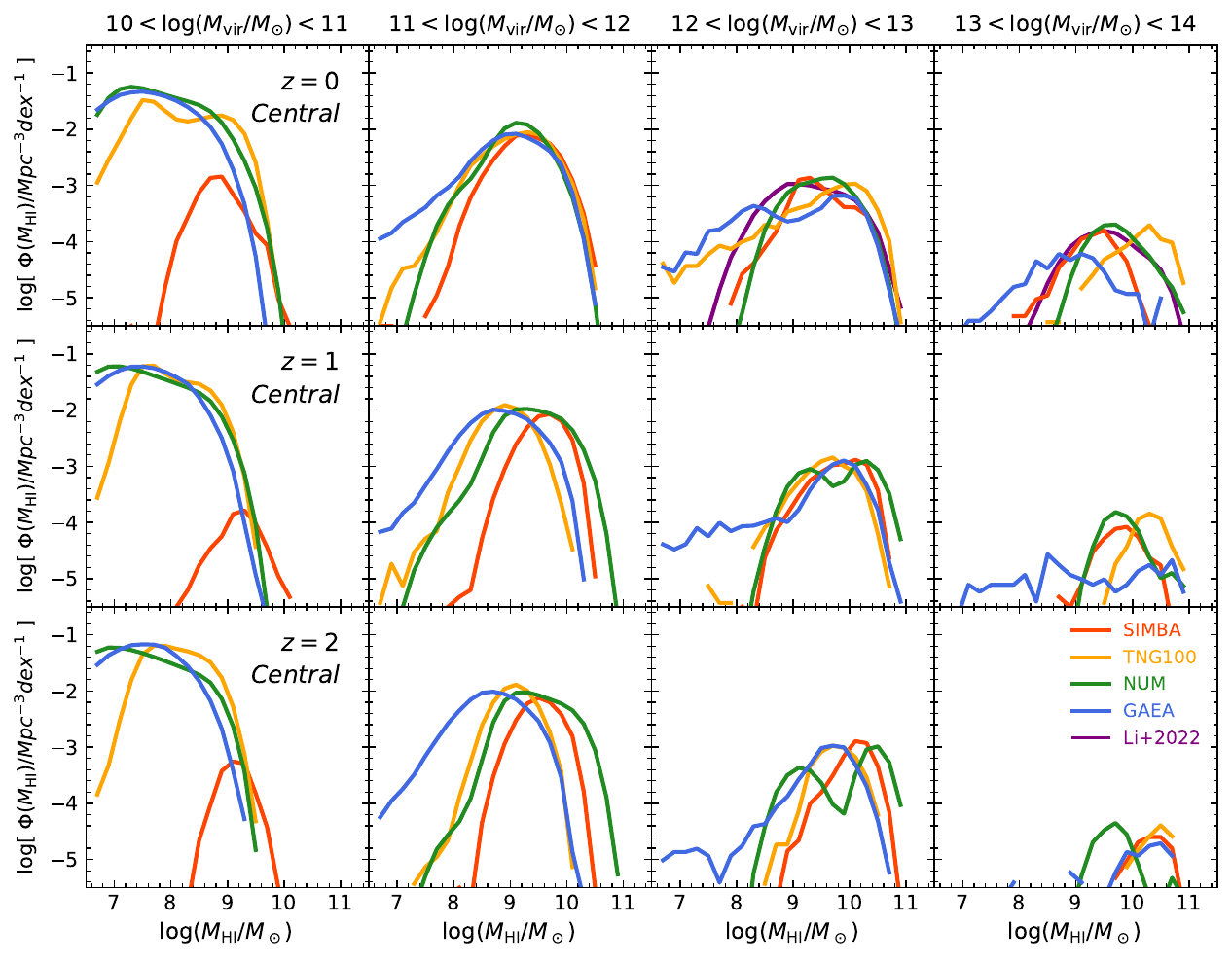}
    \caption{Comparisons between the CHIMF of central galaxies in the different simulations. The labels are the same as in Figure \ref{fig_HIMF_all}. The results of \cite{Li_2022} are plotted in purple solid lines.  {Within the mass interval $12 < \log(M_{\rm vir}/\msun) < 14$, the count of halos in GAEA is similar to that in SIMBA, whereas TNG has only about 50\% as many as SIMBA.} 
    %   {Given the rarity of massive halos and its potential impact on CHIMF, we present total host halo counts across different models of a given panel, distinguished by corresponding colours, for equitable comparison.}
    }
    \label{fig_CHIMF_cen}
\end{figure*}

\begin{figure*}
    \centering
    \includegraphics[width=0.73\textwidth]{ 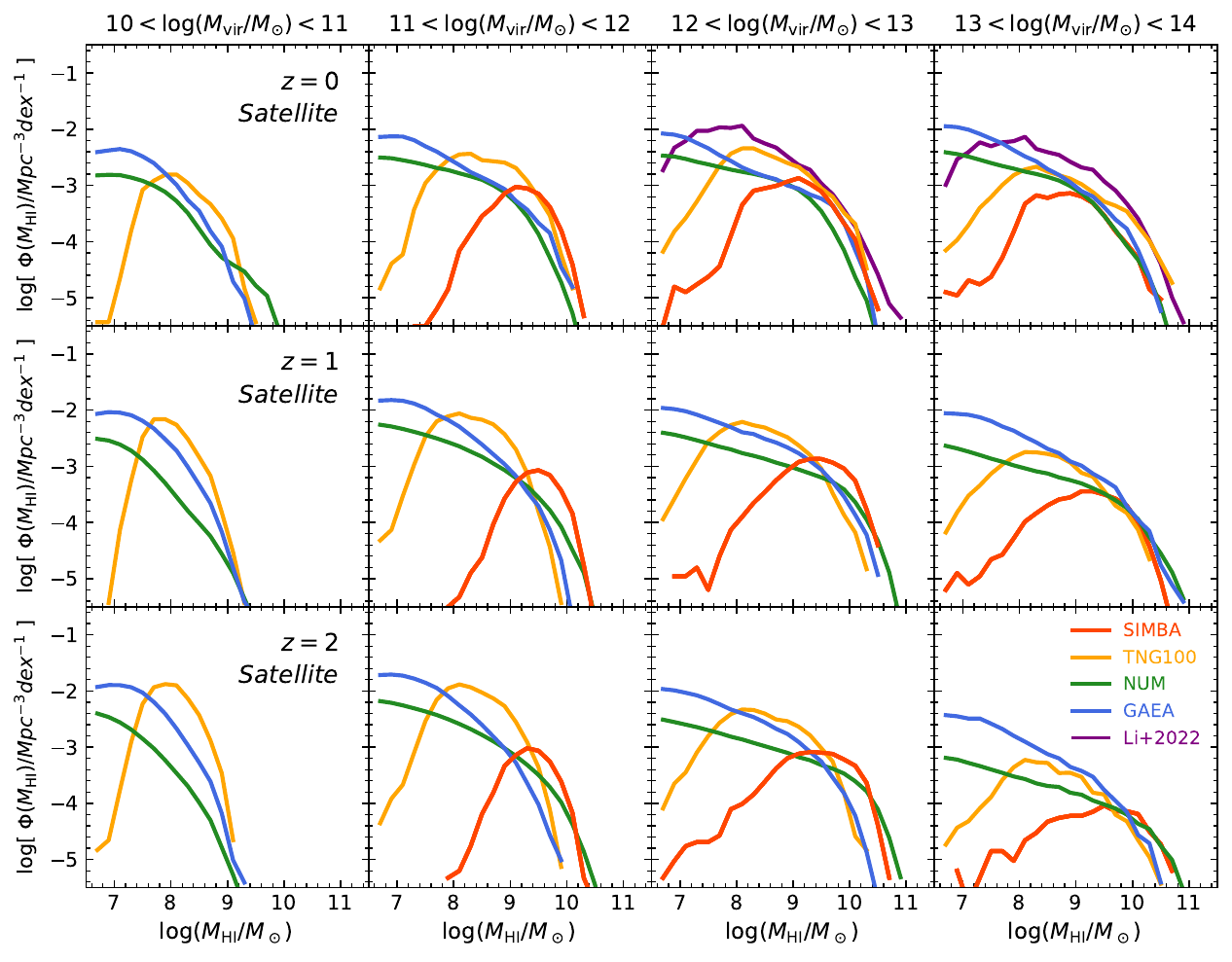}
    \caption{Comparisons between the CHIMF of satellite galaxies in the different simulations. The labels are the same as in Figure \ref{fig_HIMF_all}. Results from \cite{Li_2022} are plotted in purple solid lines.   {Within the mass interval $12 < \log(M_{\rm vir}/\msun) < 14$, the count of halos in GAEA is similar to that in SIMBA, whereas TNG has only about 50\% as many as SIMBA.} 
    %   {Given the rarity of massive halos and its potential impact on CHIMF, we present total host halo counts across different models of a given panel, distinguished by corresponding colours, for equitable comparison.}
    }
    \label{fig_CHIMF_sat}
\end{figure*}

\subsection{Conditional \hi Mass Function}\label{subsec_CHIMF}

To further investigate the cause of the differences, we calculated the HIMF in halo mass bins ranging from $10^{10}\msun$ to $10^{14}\msun$, which is called Conditional HIMF (CHIMF).   {As the same as in Section~\ref{subsec_HIMF}, we use all galaxies within the given mass bins to compute the CHIMFs, applying resolution limits of $M_\ast > 10^6 M_\odot$ for fair comparison.}

We include an additional model at $z=0$ from \cite{Li_2022} who investigated the CHIMF based on an \hi mass estimator that fully utilises the observed galaxy properties. 
  {Their \hi mass estimator is formulated as the \hi-to-stellar mass ratio ($\log (M_\hj/M_\ast)$), utilizing a linear combination of four galaxy parameters: surface stellar mass density ($\log \mu_\ast$), color index ($\rm u - r$), stellar mass ($\log M_\ast$), and concentration index ($\log (R_{90}/R_{50})$). Assuming a Gaussian error distribution, they fit this estimator to the xGASS sample, considering an \hi mass threshold of $\log (M_\hj/M_\odot) > 9.5$ due to potential nondetections and incompleteness at the low-mass end.} However, they only presented the CHIMF for halos of $M_{\rm vir}>10^{12}\msun$, which are shown as the purple lines in Figure~\ref{fig_CHIMF_cen}. 

Figure~\ref{fig_CHIMF_cen} displays the CHIMF of central galaxies for various simulations, represented by solid lines of different colours. Each column represents a different range of host halo mass, while each row represents a different redshift. There is a significant decrease in the peaks of the CHIMF as the host halo increases in all redshift ranges, indicating fewer \hi-rich galaxies in more massive halos. This indicates that most \hi in the universe is hosted by central galaxies in low-mass halos of $M_{\rm vir} < 10^{12}\msun$, as also clearly shown in Figure~19 of \cite{guo2023NUM}. The HIMF for $M_\hj<10^9\msun$ is apparently dominated by central galaxies in halos of $M_{\rm vir} \sim 10^{10}\text{-}10^{11}\msun$, while the high-mass end is mainly contributed by central galaxies with $M_{\rm vir} \sim 10^{11}\text{-}10^{12}\msun$. As the redshift decreases to 0, the HIMF increases in the highest halo mass range ($M_{\rm vir} \sim 10^{13}\text{-}10^{14}\msun$). These massive halos typically have a late formation time and higher probabilities of accumulating \hi gas through mergers \citep{guo2023NUM}. 

In general, different theoretical models show similar distributions of CHIMFs for central galaxies at $z=0$, but the   {differences} become larger at higher redshifts as in Figure~\ref{fig_HIMF_all}. We note that there is a bimodal distribution in the CHIMF of the central galaxies in the NUM model, particularly in the halo mass range of $M_{\rm vir} \sim 10^{12}\text{-}10^{13}\msun$   {at $z>1$}. It is mainly caused by the bimodal distribution of SFRs in different halo mass bins in the \textsc{UniverseMachine} model \citep{Behroozi_2019}. This also aligns with the result of \cite{Li_2022} where they found that the CHIMFs of red central galaxies peak at lower $M_\hj$ than those of the blue centrals (  {their Fig.~7}). Interestingly, GAEA tends to predict a much higher \hi abundance for low-\hi galaxies in massive halos, which is not seen in other models except for the case of TNG100 in the halo mass bin of $12<\log(M_{\rm vir}/\msun)<13$ at $z=0$.   

Figure \ref{fig_CHIMF_sat} shows the CHIMF of satellite galaxies in different redshift and halo mass bins, as in Figure \ref{fig_CHIMF_cen}. Similarly to central galaxies, the massive end of the satellite HIMF is mostly contributed by galaxies in massive halos. Different models have similar shapes of satellite CHIMF at the massive end, but they differ significantly for low-\hi galaxies. However, since the total HIMF is dominated by central galaxies, these differences in the satellite CHIMF do not have any strong effect on the total HIMF and are thus hard to discriminate in   {observations. }

\begin{figure*}
    \centering
    \includegraphics[width=0.77\textwidth]{ 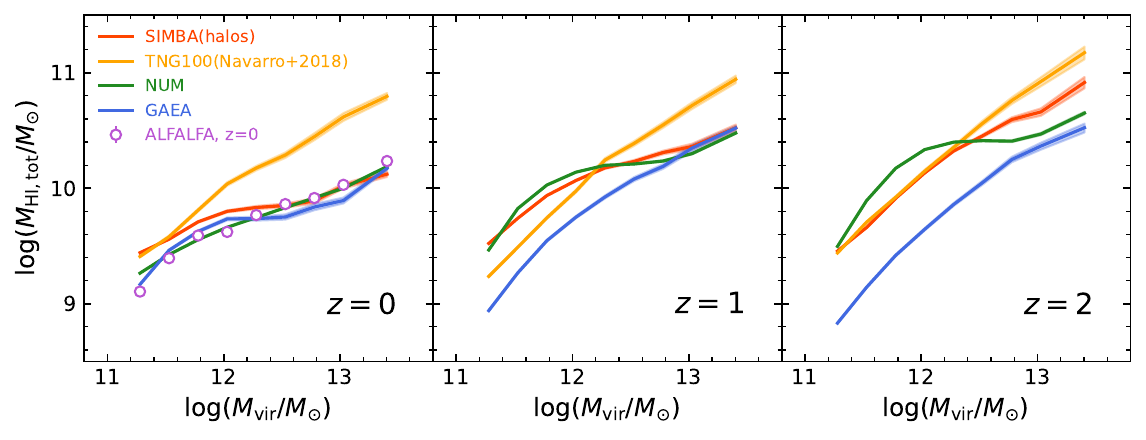}
    \caption{Comparisons between total \hi mass and halo mass in the different simulations and the observations. Solid lines and dots represent the models and the observations, respectively. The shadow area represents the $1\sigma$ error range calculated with bootstrap. Different columns represent different redshifts. Measurements of the \hi-halo mass relation at $z=0$ with confusion correction in \cite{Guo_2020} are shown as open purple circles.   {We include all \hi in galaxies and IGM as in Figure~\ref{fig_OmegaHI}, so we employ the results of \cite{Navarro_2018} for TNG and \hi in halos for SIMBA.  }}
    \label{fig_MHIMH}
\end{figure*}

\begin{figure*}
    \centering
    \includegraphics[width=0.77\textwidth]{ 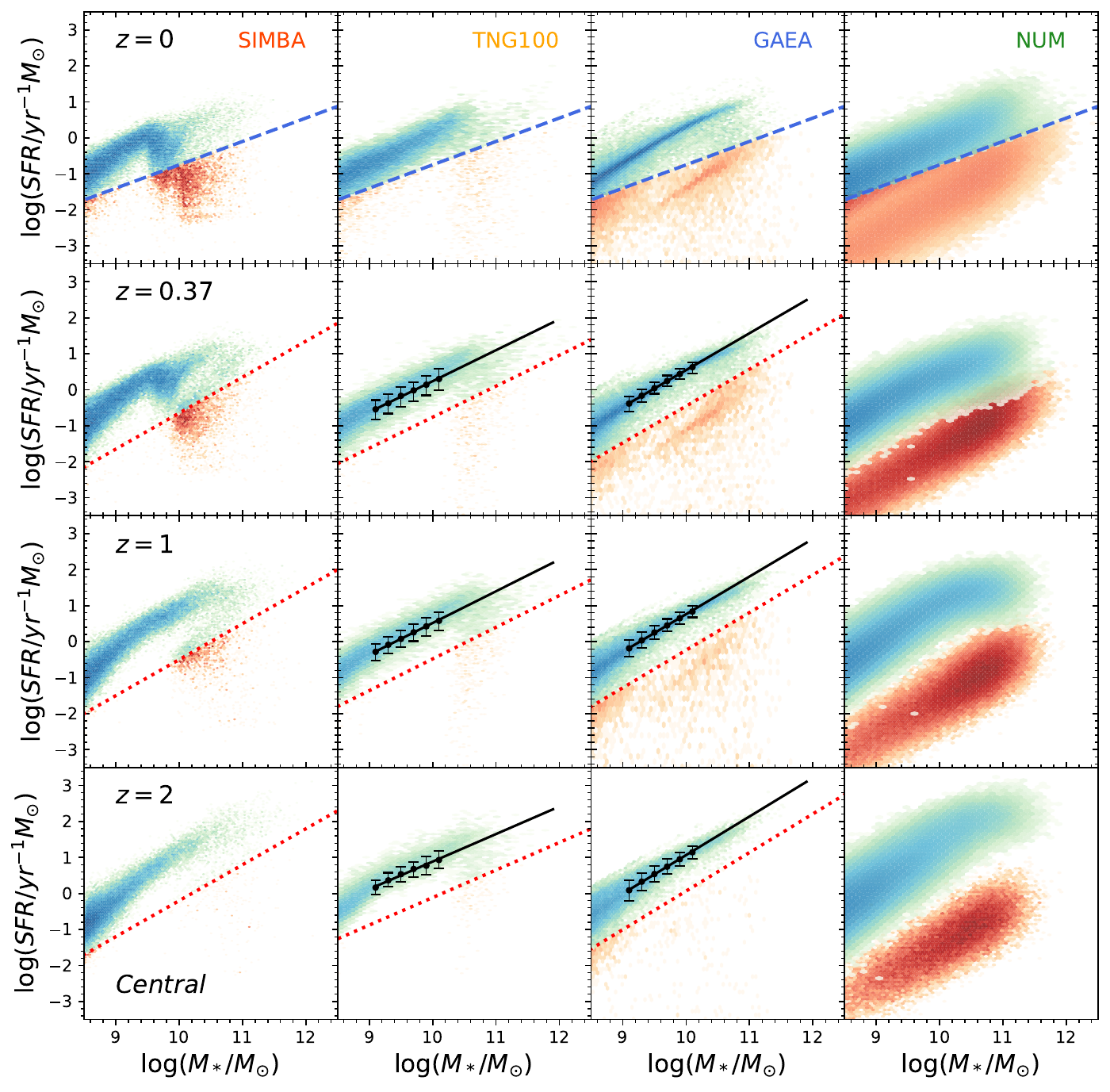}
    \caption{  {Stellar mass and SFR relation of different simulations in variant redshift. The blue scatters represent the SFGs, and the red scatters represent the QGs. The top row shows the relation at z = 0, and the SFR cuts are plotted in blue dashed lines. We use the same SFR cut described in Equation~\ref{eq0} for equal comparison. The rows below show the relation at z = 0.37, 1, and 2, respectively. For SIMBA, TNG and GAEA, the SFR cuts follow the Equation~\ref{eq1}\text{-}\ref{eq3}, shown in red dotted lines. As for NUM, we follow the same definition of SFMS as in Eq.~5 of \cite{guo2023NUM} and use the cut as 1~dex below the SFMS. }}
    \label{fig_SMSFR}
\end{figure*}

\subsection{\hj-halo Mass Relation}\label{subsec-HIHM}
The \hi-halo mass relation, denoted as $\langle M_{\text{\hi},\rm tot}|M_{\rm vir}\rangle$, characterises the total mass of the \hi gas (including all member galaxies   {and the IGM}) in halos of varying virial masses   {($M_{\rm vir}$)}. \cite{Guo_2020} measured the \hi-halo mass relation at $z\sim0$ by selecting halos from the overlapping areas of ALFALFA and galaxy groups constructed from SDSS DR7 \citep{Lim2017}. Figure~\ref{fig_MHIMH} shows the total \hi-halo mass relation for different simulations and the observations, represented by solid lines and dots, respectively. Each column corresponds to a different redshift. Measurements at $z=0$ from \cite{Guo_2020} are represented by open purple circles with error bars.

Except for TNG100, which overpredicts the total \hi mass in massive halos, all other models show reasonable agreement with the observation at $z=0$. The NUM model fits the \hi-halo mass relation well by construction. SIMBA slightly overpredicts $M_{\hj,{\rm tot}}$ in low-mass halos by around 0.3~dex. As discussed in \cite{Ma2022}, the overabundance of \hi gas for massive halos in TNG100 is mainly caused by the kinetic AGN feedback in massive galaxies that only redistributes the cold gas from the inner stellar disk to the outer region without enough depletion.

The differences among the models become much larger at higher redshifts. The NUM model agrees better with SIMBA at $z=1$ and $z=2$, where they both predict much more \hi gas at high redshifts compared to halos at $z=0$. There is not much evolution in the \hi-halo mass relation of TNG100, with only slightly higher \hi masses in the most massive halos at higher redshifts. GAEA shows a comparable amount of \hi gas with NUM for $M_{\rm vir}>10^{13}\msun$ at all redshifts, but predicts a much smaller $M_{\hj,{\rm tot}}$ for low-mass halos at $z=1$ and $z=2$. As we show in Figures~\ref{fig_CHIMF_cen} and~\ref{fig_CHIMF_sat}, most \hi gas in the universe is contained in halos of $M_{\rm vir}<10^{12}\msun$.   { The lower $M_{\hj,{\rm tot}}$ at the low halo mass end leads to the lower $\Omega_\hj$ in Figure~\ref{fig_OmegaHI}.} As seen in Figure~\ref{fig_CHIMF_cen}, the CHIMF distributions for GAEA in the halo mass bin of $11<\log(M_{\rm vir}/\msun)<12$ peak at much lower $M_\hj$ values at $z>0$ compared to NUM and SIMBA, i.e., the \hi gas in these halos is significantly depleted at higher redshifts. Physically, it might be caused by the efficient consumption of \hi gas to support the high SFRs at these redshifts in GAEA and TNG100.

\begin{figure*}
    \centering
    \includegraphics[width=0.8\textwidth]{ 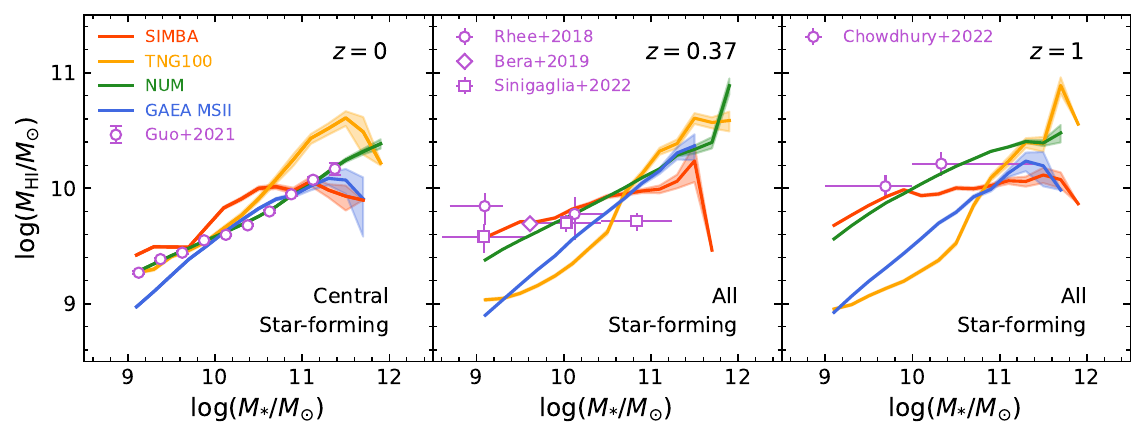}
    \caption{Comparisons between total \hi mass and stellar mass of SFGs in the different simulations and the observations. Solid lines and dots represent the models and the observations, respectively. The shadow area represents the $1\sigma$ range calculated with bootstrap. TNG100 and GAEA do not have a catalog at $z=0.37$, so we use the catalog of $z=0.5$ instead in the middle panel. Different columns represent different redshifts. The stacked \hi-stellar mass results done by \cite{Guo_2021} are shown in open purple  circles in the left panel. Measurements of \cite{Rhee_2018}, \cite{Bera_2019}, \cite{Sinigaglia_2022} at $z\sim 0.37$ are plotted as diamond, triangle, and square in the middle panel, respectively. Results of \cite{Chowdhury_2022} generated from GMRT at $z\sim 1$ are presented in open purple circles in the right panel.}
    \label{fig_MHIMS}
\end{figure*}

\begin{figure}
    \centering
    \includegraphics[width=0.4\textwidth]{ 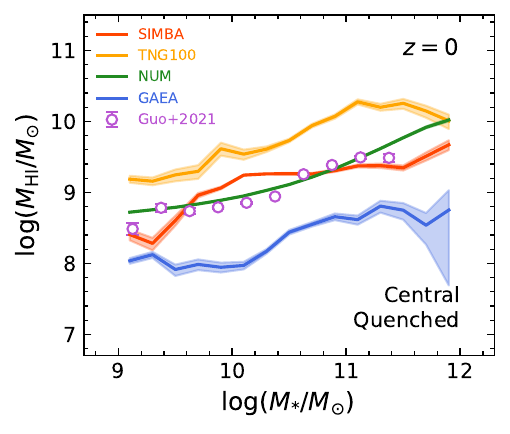}
    \caption{  {Comparisons between total \hi mass and stellar mass of QGs in the different simulations and the observations. Solid lines represents the result of models and the shadow area shows the $1\sigma$ range calculated with bootstrap. The stacked \hi-stellar mass results done by  \cite{Guo_2021} are shown in open purple  circles with errorbars.}}
    \label{fig_MHIMS_QGs}
\end{figure}

\subsection{\hj-Stellar Mass Relation}\label{subsec_HISM}
The \hi-stellar mass relation, denoted as $\langle M_{\text{\hi} }|M_\ast\rangle$, is another important relation that has been commonly measured in the literature \citep[see e.g.,][]{Saintonge2022}. Using the \hi spectra stacking technique, \cite{Guo_2021} accurately measured $\langle M_{\text{\hi} }|M_\ast\rangle$ for star-forming and quenched \emph{central} galaxies (categorised based on SFR) using the same galaxy sample as in \cite{Guo_2020}. This is particularly crucial for quenched galaxies (QGs), as most of them lack individual 21~cm detections \citep{Catinella2018}. We compare their measurements with model predictions at $z=0$ and the \hi-stellar mass measurements from \cite{Chowdhury_2022} obtained with GMRT at $z\sim 1$. At $z\sim 0.37$, we adopt the measurements of \cite{Rhee_2018} and \cite{Bera_2019} from GMRT as well as the measurements of \cite{Sinigaglia_2022} from MeerKAT. We focus on the comparisons for star-forming galaxies as they are the main targets of observation at high redshifts. 

All these observations adopt the \hi stacking method to obtain the average \hi masses for star-forming galaxy samples, that is, $\langle M_\hj\rangle$. We measure the corresponding \hi masses in the same way in different models,   {as $\log(\Sigma_i M_{\hj,i}/N_{{\rm halo}, i})$, where $N_{{\rm halo},i}$ is the number of host halos in a given halo mass bin $i$, and the summation of $\Sigma_i M_{\hj,i}$ is over all galaxies in the halo mass bin.} To ensure fair comparisons, we only consider   {star-forming} galaxies with stellar masses in the range of $10^9-10^{12}\msun$ that can be fully resolved at different simulation resolutions.

For fair comparisons, at $z=0$ we adopt the same SFR cut in all models to select star-forming \emph{central} galaxies as in \cite{Guo_2021}:
\begin{equation}\label{eq0}
    \log({\rm SFR}_{\rm cut, z = 0}/{\rm yr}^{-1}\msun) = 0.65\log M_\ast - 7.25
\end{equation}

At high redshifts, different SFR cuts are adopted for different simulations. For SIMBA, we follow the cut to separate star-forming and QGs as in \cite{Dave2019},
\begin{equation}\label{eq1}
    \log({\rm SFR}_{\rm cut,SIMBA}/{\rm yr}^{-1}\msun) = \log M_\ast - 10.8 + 0.3z
\end{equation}

For TNG and GAEA, we first determine the SFMS by fitting a power law relation to the number density peaks of galaxies with specific SFR (sSFR) larger than $10^{-11}{\rm yr}^{-1}$, as in \cite{Ma2022}. Then the cuts between the star-forming and QGs are simply set as 1~dex below the SFMS.

The best-fit cuts at different redshifts for TNG are as follows.
\begin{eqnarray}\label{eq2}
    % \log({\rm SFR}_{\rm cut,TNG,z=0}/{\rm yr}^{-1}\msun) = 0.83\log M_\ast - 9.34\\
    \log({\rm SFR}_{\rm cut,TNG,z=0.5}/{\rm yr}^{-1}\msun) = 0.86\log M_\ast -9.38\\
    \log({\rm SFR}_{\rm cut,TNG,z=1}/{\rm yr}^{-1}\msun) = 0.88\log M_\ast -9.30
    % \log({\rm SFR}_{\rm cut,TNG,z=2}/{\rm yr}^{-1}\msun) = 0.77\log M_\ast -7.77
\end{eqnarray}

Similarly, the cuts for GAEA galaxies are,
\begin{eqnarray}\label{eq3}
    % \log({\rm SFR}_{\rm cut,GAEA,z=0}/{\rm yr}^{-1}\msun) = 1.01\log M_\ast -10.83\\
    \log({\rm SFR}_{\rm cut,GAEA,z=0.5}/{\rm yr}^{-1}\msun) = 1.02\log M_\ast -10.69\\
    \log({\rm SFR}_{\rm cut,GAEA,z=1}/{\rm yr}^{-1}\msun) = 1.05\log M_\ast -10.69
    % \log({\rm SFR}_{\rm cut,GAEA,z=2}/{\rm yr}^{-1}\msun) = 1.07\log M_\ast -10.65
\end{eqnarray}

For NUM galaxies, we follow the same definition of SFMS as in Eq.~5 of \cite{guo2023NUM} and use the cut as 1~dex below the SFMS. The above cuts do not have a significant impact on our conclusions. All SFR cuts are plotted in red dotted lines (blue dashed lines at z = 0 ) in Figure~\ref{fig_SMSFR}, the blue scatters represent the SFGs, and the red scatters represent the QGs.
% %%%%%%%%%%

Comparisons between the \hi-stellar mass in the different simulations and the observations for star-forming galaxies are shown in Figure \ref{fig_MHIMS}. In this figure, the solid lines and dots represent the models and the observations, respectively. The shaded area represents the $1\sigma$ error in the model predictions. Each column in the figure corresponds to a different redshift. The left panel of Figure~\ref{fig_MHIMS} shows the comparisons of different models. As in the case of \hi-stellar mass relation, SIMBA and GAEA agree reasonably well with the measurement of \cite{Guo_2021}. GAEA tends to slightly underestimate the \hi masses in low-mass central galaxies of $M_\ast<10^{10}\msun$. This corresponds to the underestimates of HIMF in the mass range of $8.5<\log(M_\hj/\msun)<9.5$ seen in Figure~\ref{fig_HIMF_all}. Similarly, TNG100 overestimates the \hi masses for massive galaxies (as also shown in Fig.~2 of \citealt{Ma2022}) which causes the overestimates of \hi-rich galaxies in the massive end of the HIMF.

We note that the TNG100 and GAEA simulations do not have the output of \hi catalogues at $z=0.37$, so we use the catalogues of $z=0.5$ in the middle panel instead. The stacked \hi measurements of \cite{Rhee_2018}, \cite{Bera_2019}, and \cite{Sinigaglia_2022} at $z\sim0.37$ are shown as different symbols. Measurements of \cite{Chowdhury_2022} at $z\sim1$ are shown as open purple circles in the right panel. The   {differences} among the different models become significantly larger at these higher redshifts. The general trend is consistent with the results in the \hi-halo mass relation. TNG100 and GAEA tend to underestimate the \hi content in low-mass galaxies, and their redshift evolution of the \hi-stellar mass relation is   {mild}.   
Currently, our ability to observe \hi targets at high redshifts is very limited, and most of the available data are around $z=0.37$.   {The large   {differences} between models at higher redshift indicate that HI surveys at higher redshift are essential on constraining galaxy formation models.}
% As a result, the accuracy of existing models for even higher redshifts is not well constrained by observations and will be improved with future \hi surveys.

  {In addition, we compare the \hi - stellar mass relation for QGs at $z=0$ in Figure~\ref{fig_MHIMS_QGs}. The results for different models are plotted in solid lines with same colours as in Figure~\ref{fig_MHIMS} and are compared to the stacked \hi observations of \cite{Guo_2021} (purple circles). We found that TNG tends to overestimate the \hi in QGs and GAEA tends to underestimate them, while SIMBA fit the observations quite well. }

\begin{figure*}
    \centering
    \includegraphics[width=0.8\textwidth]{ 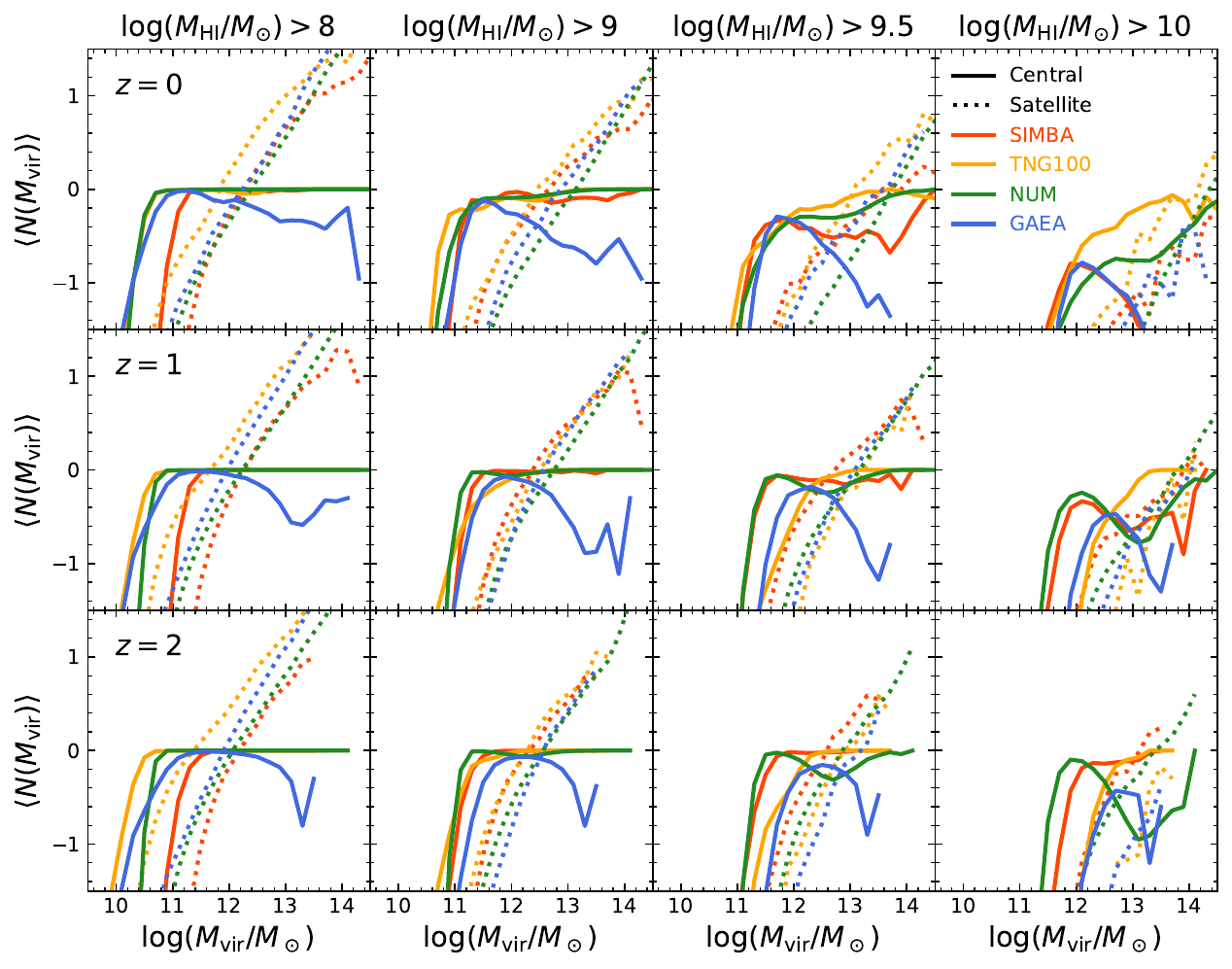}
    \caption{Comparisons of HOD in the different simulations,   {solid lines and dotted lines represent the central and satellite galaxies}, respectively. Different columns represent different \hi mass bins, and different rows represent different redshifts.}
    \label{fig_HOD}
\end{figure*}

%\begin{figure*}
%    \centering
%    \includegraphics[width=0.8\textwidth]{ HI-HOD-N_M-GUO-QIN-z0.pdf}
%    \caption{Comparisons of HOD between NUM and \cite{Qin_2022} at z$\sim$0,   {solid lines and dashed lines represent the central and satellite galaxies}, respectively, and orange and green lines represent the data of \cite{Qin_2022} and NUM, respectively.   {The shaded region shows the 1$\sigma$ error range.} Different panels represent different \hi mass bins.}
%    \label{fig_HOD_QinFei}
%\end{figure*}

\subsection{Halo Occupation Distribution}\label{subsec_HOD}
To better understand the different \hi distributions in the models, it is also helpful to compare the halo occupation distributions for galaxies with different \hi mass thresholds \citep{Guo2017}. The Halo Occupation Distribution (HOD) characterises the likelihood distribution $P(N|M_{\rm vir})$ for the presence of $N$ galaxies within a specific dark matter halo with virial mass $M_{\rm vir}$ \citep[see e.g.,][]{Jing1998,Berlind2002,Zheng2005,Guo2015}.
In principle, the information content of the HOD is the same as that of the CHIMF \citep{Yang2003}. But the HOD provides a straightforward way to understand the distributions of central and satellite galaxies in different halos \citep[see e.g.,][]{Qin_2022}. 

Figure \ref{fig_HOD} illustrates the HOD of the various models,   {with solid lines for central galaxies and dotted lines for satellite galaxies.} Each column corresponds to a different \hi mass threshold sample as labelled, and each row represents a different redshift. The satellite HODs for galaxies with different $M_\hj$ thresholds have similar shapes in different models, but with slightly different amplitudes. TNG100 generally predicts more \hi-rich satellites than other models, especially for the samples of $M_\hj>10^8\msun$. This means that the \hi depletion for satellites in TNG100 is not efficient enough for low-mass galaxies. 

TNG100, SIMBA and NUM have similar central HODs for low-$M_\hj$ galaxies. We note that for the sample of $M_\hj>10^8\msun$, SIMBA seems to predict a much higher central halo mass cutoff than other models. But in fact it is just caused by the low-mass resolution of the SIMBA simulation. As seen in Figure~\ref{fig_HIMF_all}, there are very few galaxies with $M_\hj<10^9\msun$ in SIMBA. Therefore, the HODs of $M_\hj>10^8\msun$ and $M_\hj>10^9\msun$ are almost the same in SIMBA because of the lack of low-$M_\hj$ galaxies. 

The central HOD in GAEA is approaching unity in halos of $M_{\rm vir}\sim10^{11}\text{-}10^{12}\msun$, but begins to decrease for massive halos. This indicates that central galaxies in more massive halos are increasingly \hi-poor in GAEA. The \hi depletion in these central galaxies is more efficient than in other models. This trend is even more prominent in samples with higher \hi thresholds. As the \hi-stellar mass relation for star-forming central galaxies in GAEA is consistent with that of NUM, it implies that the \hi depletion in QGs of GAEA is likely too strong. We confirm this by comparing the stacked \hi-stellar mass relation for quenched centrals in \cite{Guo_2021} with the corresponding predictions from GAEA and find a 0.7~dex lower $M_\hj$ for quenched centrals in GAEA  {, as shown in Figure~\ref{fig_MHIMS_QGs}}. These differences will be further investigated in future work.

\section{Discussion}\label{sec_dicussion}
The statistical measurements of \hi observations are still very limited, especially beyond the local universe, due to the sensitivity limits of radio telescopes. It is therefore very important to have theoretical models that can help understand the evolution history of the cosmic \hi content. The motivation of the NUM model is to build an empirical model to successfully describe the available \hi measurements without involving complicated baryonic physical processes. Comparisons between NUM and other theoretical models would enhance our understanding of the properties of galaxy \hi gas. 

The HIMF is commonly used to calibrate \hi distributions in theoretical models, as it is well measured with blind \hi surveys and robust against observational systematics. For example, the GAEA model is calibrated with the HIMF of \cite{Haynes_2011} using the early data release of ALFALFA, which is consistent with the latest HIMF measurement of \cite{guo2023NUM}. Interestingly, neither TNG100 nor SIMBA is tuned to match the HIMF at $z\sim0$, but they show a general agreement with the observation. This might be related to the fact that these simulations are tuned to reproduce the galaxy stellar mass functions, stellar-halo mass relations, and SFR distributions at $z\sim0$. As discussed in \cite{guo2023NUM}, the galaxy mass \hi is mainly determined by the halo mass, the halo formation time, and the SFR. Therefore, the HIMF is reasonably reproduced.

However, HIMF by itself is insufficient to completely distinguish between models, as it only describes the number densities of galaxies with differing \hi masses. The \hi-halo and \hi-stellar mass relations are crucial for   {connecting} the \hi gas and its host galaxies (halos). As depicted in Figure~\ref{fig_HIMF_all}, the four theoretical models align reasonably well with the observed HIMF at $z\sim0$. TNG100 tends to slightly overestimate the HIMF at the higher mass end, whereas GAEA underestimates it in the mass range of $8.5<\log(M_\hj/\msun)<9.5$. By analysing the \hi-stellar mass relations of various models at $z=0$, it is evident that these   {differences} arise from the inconsistency with the observed \hi-stellar mass relation. This inconsistency is also seen in the HOD distributions, where TNG100 has notably higher central HODs, and GAEA demonstrates lower central HODs for \hi-rich galaxies in massive halos (Figure~\ref{fig_HOD}).

As shown in \cite{Ma2022}, the main driver for the different \hi distributions in the theoretical models is the AGN feedback mechanism. By comparing the stacked \hi masses between AGNs and star-forming galaxies with the same $M_\ast$ and SFR, \cite{Guo2022} found that the \hi depletion from the AGN feedback is much stronger for galaxies with higher SFRs and higher AGN luminosity. \cite{Dave2020} also compared the cold gas distributions in the simulations of TNG100, SIMBA, and EAGLE. They found the same trend of differences in the HIMF as in this study. They concluded that the thermal-based AGN feedback in EAGLE by raising the temperature of gas and the kinetic AGN feedback in SIMBA and TNG will affect the HIMF in different ways. 

The \hi-halo mass relation provides more insight to understand the effect of AGN feedback. \cite{Baugh2019} found a dramatic drop in total \hi mass in halos of $11<\log(M_{\rm vir}/\msun)<12$ due to the strong suppression of gas cooling from efficient AGN feedback in the GALFORM semi-analytical model. \cite{Chauhan2020} also found a similar level of drop in the \hi-halo mass relation at a slightly higher halo mass range in the SHARK semi-analytical model. However, they attributed this drop to the missing contribution of the \hi gas in the circumgalatic medium. As we show in Figure~\ref{fig_MHIMH}, such a strong drop is not seen in the hydrodynamical simulations of TNG100 and SIMBA, as well as the semi-analytical model of GAEA \citep[see also Fig.~5 of][]{Spinelli_2020}. Therefore, the \hi-halo mass relation can be used to discriminate between the different AGN feedback mechanisms. 

The CHIMF and HOD include more detailed information on the distributions of \hi gas in central and satellite galaxies. \cite{Li_2022} investigated the CHIMF for the red and blue centrals and found that the CHIMF distributions of the blue centrals are narrower and peak at higher $M_\hj$ values. This is consistent with the NUM model of the SFR dependence of the \hi mass. It is also manifested in the central HOD distributions of different threshold samples $M_\hj$ in Figure~\ref{fig_HOD}, which makes CHIMF and HOD useful statistics for understanding the \hi-halo
connection. 

\section{Conclusions}\label{sec_conclusion}
Hydrodynamical and semi-analytical models offer a convenient approach to studying the universe. However, both models are dependent on physical processes and face challenges when it comes to predicting the properties of \hi. To address this issue, we employ the empirical model of NUM \citep{guo2023NUM}, which is constructed based on observations and demonstrates excellent predictive capabilities for \hi statistics. In this study, we compare NUM with other models in terms of the CHIMF, the \hi-halo and \hi-stellar mass relations, and the HOD distributions for \hi gas. Our main conclusions are summarised as follows.

(i) All the theoretical models of TNG100, SIMBA, GAEA and NUM show reasonable agreement with the observed HIMF at $z\sim0$. TNG100 slightly overestimates the massive end of HIMF and GAEA underestimates HIMF in the range of $M_\hj\sim10^{8.5}\text{-}10^{9.5}\msun$. The   {differences} in the models become much larger at higher redshifts of $z=1$ and $z=2$. The NUM model still presents remarkable agreement with the HIMF measurements of \cite{Chowdhury2024} at $z=1$, but other models tend to underestimate the high-mass end of the HIMF at this redshift. The evolutionary trend at the high-mass end is even stronger at $z=2$, with NUM predicting much more \hi-rich galaxies, SIMBA showing roughly constant number densities, and both TNG100 and GAEA indicating much lower amplitudes.  

(ii) From the comparisons for the CHIMF, it is clear that the HIMF is mainly dominated by central galaxies at all \hi masses and redshifts. The low-mass end of the HIMF ($M_\hj<10^{9}\msun$) is mainly contributed by halos with $10 < \log (M_{\rm vir}/\msun) < 11$, while the high-mass end is dominated by halos with $11 < \log (M_{\rm vir}/\msun) < 12$. As in the case of HIMF, the differences in the CHIMF distributions of the models are generally small for central galaxies at $z\sim0$, but become larger at higher redshifts, especially for galaxies in massive halos. The differences in the satellite CHIMFs are even greater. Future \hi surveys of satellite galaxies will provide tighter constraints on different models.
    
(iii) The   {differences} in the HIMFs of different models can be better understood using the CHIMF and the \hi HOD. It is evident in the central \hi HOD distributions that TNG100 overpredicts the number of central galaxies with high $M_\hj$ in massive halos and GAEA shows very strong depletion of \hi gas in quenched central galaxies of massive halos. The most likely cause of the differences is the AGN feedback mechanisms implemented in these models, as the main differences lie in the quenched populations in massive halos. 
    
\begin{acknowledgements}
This work is supported by the National SKA Program of China (grant No. 2020SKA0110100), the CAS Project for Young Scientists in Basic Research (No. YSBR-092), the science research grants from the China Manned Space Project with NOs. CMS-CSST-2021-A02 and GHfund C(202407031909). We acknowledge the use of the High Performance Computing Resource in the Core Facility for Advanced Research Computing at the Shanghai Astronomical Observatory.
\end{acknowledgements}

\bibliographystyle{aa}
\bibliography{ref}
	
\end{document}